**Electron mobility in AlN from first principles**

Amanda Wang[1], Nick Pant[1,2], Woncheol Lee[3], Jie-Cheng Chen[4,5], Feliciano Giustino[4,5], Emmanouil Kioupakis[1]

[1] Department of Materials Science and Engineering, University of Michigan, Ann Arbor, Michigan 48109, USA

[2] Applied Physics Program, University of Michigan, Ann Arbor, Michigan 48109, USA

[3] Department of Electrical and Computer Engineering, University of Michigan, Ann Arbor, Michigan 48109, USA

[4] Oden Institute for Computational Engineering and Sciences, The University of Texas at Austin, Austin, Texas 78712, USA

[5] Department of Physics, The University of Texas at Austin, Austin, Texas 78712, USA

Aluminum nitride is a promising ultra-wide band gap semiconductor for optoelectronics and power electronics. However, its practical applications have been limited by challenges with doping and achieving high electrical conductivity. Recent advances in crystal quality and defect control have led to improvements in experimentally measured mobilities. In this work, we apply first-principles calculations to determine the upper limits of the electron mobility in AlN as a function of temperature, doping, and crystallographic orientation. We account for the combined effects of electron scattering by phonons and ionized impurity to model doped systems, and examine both full and partial ionization conditions. Our results show that the piezoelectric interaction from the long-range component of the acoustic modes is the dominant source of electron-phonon scattering at room temperature. Ionized-impurity scattering starts to dominate scattering at dopant concentrations above $10^{16}$ cm$^{-3}$, reducing the mobility by more than an order of magnitude in the high doping regime. Our calculated Hall mobility values are in good agreement with experimental data for samples with comparable dopant concentrations. We also find that electron mobilities as high as 956 cm$^2$/V·s could be achievable at lower dopant concentrations.

Wurtzite aluminum nitride (AlN) is a promising semiconductor material for deep UV optoelectronics as well as high-power and high-frequency devices due to its ultra-wide band gap of 6.2 eV.[1] The super-linear scaling of the breakdown field of materials with respect to their band gap predicts that AlN has an exceptionally high breakdown field (15 MV/cm)[2], making it particularly suitable for high-power electronics. However, challenges with doping and achieving high electrical conductivity in AlN have largely limited its role to that of an electrically insulating (but thermally conducting) substrate in electronic devices.[3,4] Alternatively, alloying AlN with GaN improves the dopability, but also reduces the band gap and introduces alloy disorder, which scatters electrons and phonons and thus suppresses the mobility and thermal conductivity.[5]

It has been difficult to achieve high free electron densities and high dopant ionization ratios in AlN.[6] *n*-type dopants in AlN, the most common being Si and Ge, form negatively charged DX$^-$ centers that not only trap one extra electron instead of donating an electron, i.e., decreasing instead of increasing the free electron concentration, but also increase electron scattering through the long-range Coulomb interaction with their trapped charge.[7] While electron densities higher than $10^{15}$ cm$^{-3}$ have been reported in homojunction AlN diodes,[8] highly doped AlN samples have low electron mobilities because transport is limited by ionized-impurity scattering from the dopants.[9,10] In contrast, transport in the low-doping regime is limited by other point defects and dislocations.[11] Works that focus on high crystal quality and reducing the concentration of defects and dislocations to achieve conductive samples in the low doping regime have shown the best transport properties, with the highest mobility being 426 cm$^2$/V·s.[11–13] The high mobility values are particularly remarkable given the ultra-wide band gap of AlN. For comparison, ultra-wide band gap semiconductor $\beta$-Ga$_2$O$_3$ has a band gap of 4.7 eV and the highest experimental mobility reported is 180 cm$^2$/V·s, while first-principles calculations of mobility are lower than 300 cm$^2$/V·s.[14,15] As crystal quality and transport properties in AlN continue to improve, this raises the question of how high the mobility can be.

There exist theoretical studies of electron transport in AlN, primarily in the context of AlGaN alloys. Earlier works used Monte Carlo simulations to model transport in AlGaN alloys based on analytical band structures and scattering rates. The highest AlN mobility they calculate is 657 cm$^2$/V·s, corresponding to the lowest ionized-impurity concentration they tested of $10^{16}$ cm$^{-3}$.[16] A subsequent Monte Carlo simulation models transport in binary nitrides GaN, AlN, and InN using band structures computed with semi-empirical pseudopotentials and analytical scattering rates, reporting mobility in AlN to be 470 cm$^2$/V·s at a donor density of $10^{17}$ cm$^{-3}$.[17] A study of the Baliga figure of merit of AlGaN alloys uses analytical models for mobilities limited by phonon, alloy, and ionized-impurity scattering and combines the mobilities according to Matthiessen's rule. They calibrate the models to experimental data and calculate AlN mobility to be 413 cm$^2$/V·s, closely reproducing the value from Ref. 12.[18]

While analytical models have lower computational cost compared to calculating scattering rates and mobilities from first principles, they also have lower accuracy and predictive power, especially as the models need to be parameterized based off empirical data. Previous first-principles calculations of the electron mobility of AlN focused only on the effects of electron-phonon scattering and did not include the effects of dynamical quadrupoles or the effects of ionized-impurity scattering that enable comparisons to experimental mobility measurements of doped samples.[19–21] The more recent study of the electron mobility in AlGaN alloys by Datta *et al.*[22] does include the effects of ionized impurities, but only the long-range component of the polar optical and piezoelectric phonons are included through first principles while the ionized-impurity and deformation potential scattering are included through models. Overall, a comprehensive study that focuses on the phonon- and ionized-impurity-limited mobility in AlN that incorporates the electron-phonon coupling fully from first principles is still missing.

In this work, we calculate the electron mobility in AlN from first principles, accounting for electron scattering due to phonons and ionized impurities, to investigate the theoretical upper limit to the mobility values as a function of temperature, doping, and crystallographic orientation. We examine the temperature dependence of the phonon-limited mobility and analyze the contribution to the total scattering rate by the various phonon modes. By decomposing the scattering rate in terms of phonon energy and accounting for the magnitudes of the electron-phonon coupling matrix elements and phonon occupations, we find that the acoustic modes dominate electron-phonon scattering through the long-range piezoelectric interaction. Our calculations determine the total mobility, including the combined effects of both phonon and ionized-impurity scattering, at various electron densities and ionized-impurity concentrations. Our calculated electron Hall mobilities are in excellent agreement with experimental mobility data. We show that at low ionized impurity concentrations, Hall mobilities as high as 956 cm$^2$/V·s are possible, over two times what has been seen in experiment.

We apply density functional theory (DFT) and density functional perturbation theory (DFPT) to determine the electron, phonon, and electron-phonon coupling properties of AlN, which we then incorporate into the iterative Boltzmann transport equation (IBTE) to evaluate the electron mobility. We use the implementation of DFT and DFPT in Quantum ESPRESSO[23,24] and employ the local density approximation (LDA) to the exchange-correlation functional.[25,26] The experimental lattice parameters of a = 3.11 Å and c = 4.98 Å are used in this work,[27] because of the small differences from the relaxed lattice parameters of 3.07 Å and 4.92 Å, respectively. We apply quasiparticle corrections to the energies using the $G_0W_0$ method as implemented in BerkeleyGW (Fig. S1).[28,29] The frequency dependence of the dielectric function is included through the plasmon-pole model,[28] and the convergence of the quasiparticle corrections is accelerated using the static remainder approach.[30] We find that the quasiparticle corrections change the curvature of the bottom of the conduction band by 6% (Fig. S2) and that the phonon-limited mobility values differ by 5% or less between the DFT and $G_0W_0$ energies. Due to numerical convergence issues at high ionized-impurity concentrations in total mobility calculations when using $G_0W_0$ eigenvalues, we report all mobilities calculated using DFT eigenvalues in this work. The electron energies, velocities, and electron-phonon matrix elements are calculated on a coarse 8×8×8 Brillouin zone (BZ) sampling grid and then interpolated to fine sampling grids using maximally localized Wannier functions, as implemented in Wannier90 and EPW.[31–33] The accuracy of the electron-phonon matrix interpolation is improved with dipole and quadrupole corrections to the long-range component of the electron-phonon interactions. The dipole correction is calculated with the Born effective charges calculated by DFPT,[34] and the quadrupole correction is calculated with the quadrupole tensor (Table S1), calculated using ABINIT.[35,36] The interpolated quantities are used to solve the IBTE as implemented in the EPW code, with mobilities converging at fine

BZ sampling grids of 144×144×96. Further details of the calculation are provided in section S1 of the supplementary.

Table 1 Electronic properties of AlN, as calculated in this work and compared to previous calculations and experiments. Values listed in parentheses are calculated with $G_0W_0$.

|  | $E_g$ (eV) | $m^* \perp c$ | $m^* \parallel c$ | $\varepsilon_0^\perp$ | $\varepsilon_0^\parallel$ | $\varepsilon_\infty^\perp$ | $\varepsilon_\infty^\parallel$ |
|---|---|---|---|---|---|---|---|
| This work | 4.31 (6.31) | 0.33 (0.32) | 0.31 (0.33) | 8.43 | 10.11 | 4.40 | 4.62 |
| Experiment | 6.13–6.28[1] | 0.29 – 0.45[37] | | 7.76[38] | 9.32[38] | 4.16[38] | 4.35[38] |
| Previous theory | 6.31[39] | 0.34[39] | 0.32[39] | 8.5[40] | | 4.38[41] | 4.61[41] |

We calculate the temperature dependence of the phonon-limited drift mobilities and find the room-temperature mobilities to be 871 cm$^2$/V·s and 619 cm$^2$/V·s along the in-plane and out-of-plane directions, respectively. Figure 1a shows how the mobilities, calculated at an electron concentration of $10^{15}$ cm$^{-3}$, decrease with increasing temperature due to increasing phonon occupation and consequently phonon scattering. We fit the temperature dependence of the mobility to the following model, which describes a characteristic mobility at low temperatures and one at high temperatures, combined according to Matthiessen's rule:[42,43]

$$\mu_{\text{ph}}(T) = \left( \frac{1}{\mu_{\text{low}}} e^{-T_{\text{low}}/T} + \frac{1}{\mu_{\text{high}}} e^{-T_{\text{high}}/T} \right)^{-1}. \tag{1}$$

The fitted parameters are listed in Table SII. The characteristic temperatures $T_{\text{low}}$ and $T_{\text{high}}$, which indicate the energies of the dominant phonon modes at low and high temperatures, are approximately 8 and 7 meV, and 106 and 104 meV for the in-plane and out-of-plane directions, respectively. These energies correspond to that of the low-energy acoustic modes and the highest-energy longitudinal optical modes, respectively.

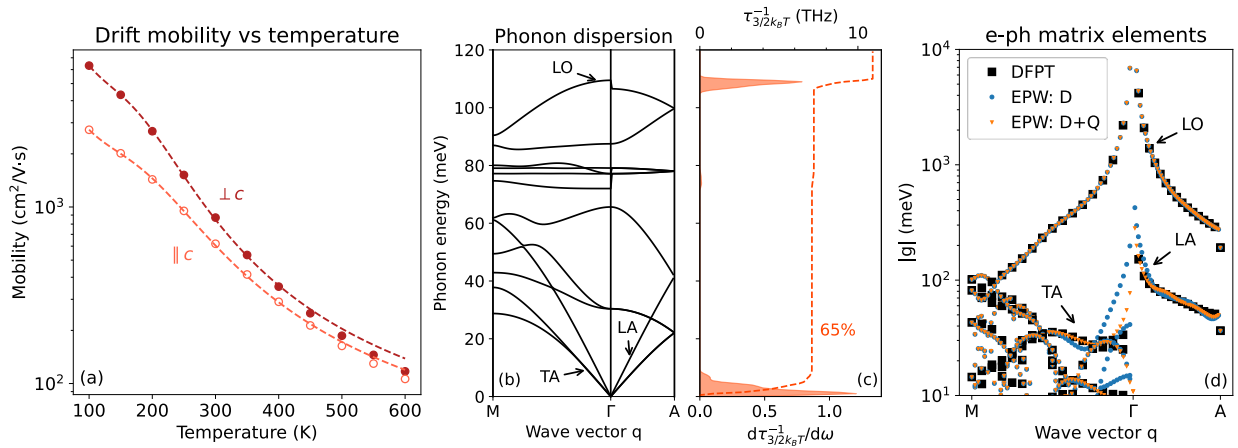

Figure 1 (a) Phonon-limited electron drift mobility of AlN along the in-plane and out-of-plane directions plotted as a function of temperature. The dashed lines are fits of the data using the function in the text. (b) Phonon dispersion of AlN along high-symmetry paths from M to $\Gamma$ (in-plane) and from $\Gamma$ to A (out-of-plane). (c) Distribution of the room-temperature scattering rate for a representative electron state at $3/2 k_B T$ above the band edge as a function of phonon energy. The peaks indicate the contribution to electron scattering by phonons at each energy, while the dashed line shows the cumulative contribution to the total scattering rate. Scattering by the low-energy, long-range acoustic modes (piezoelectric scattering) contribute to 65% of the total scattering rate at room temperature. (d) Electron-phonon matrix elements along the same high-symmetry paths as the phonon dispersion. The values directly calculated by DFPT are shown in black squares, and the values interpolated by EPW with just the dipole correction and the dipole + quadrupole corrections are shown in blue circles and orange triangles, respectively. Including both dipole and quadrupole corrections significantly improves the accuracy the interpolation near $\Gamma$. The longitudinal optical mode has the largest matrix elements, followed by the longitudinal and transverse acoustic modes.

By analyzing the contributions of the different phonon modes to the electron scattering rate, we further confirm that the acoustic modes and the highest-energy longitudinal (polar) optical modes dominate electron scattering by phonons. This can be seen in Figures 1b and 1c, which show the phonon dispersion alongside the breakdown of the room-temperature electron scattering rate by phonon energy. There are two predominant peaks in the breakdown of the scattering rate, one in the range of 0-10 meV and the other at 105-110 meV, which agree with the fitted characteristic temperatures discussed in the previous paragraph. The low energy of the first peak shows that long-wavelength acoustic phonons near $\Gamma$ ($q \to 0$) primarily contribute to scattering, pointing to the long-range piezoelectric interaction dominating electron-phonon scattering. Examining the electron-phonon coupling matrix elements (Figure 1d), which represent the strength of electron-phonon scattering, also shows that the longitudinal acoustic and optical modes are the most dominant, followed by the transverse acoustic. The strength of the contribution of the acoustic modes, which contribute to 65% of the scattering rate at room temperature, is unusual for a polar semiconductor in which the polar optical modes are typically the dominant source of scattering.[44,45] However, in AlN (and similarly in GaN) the highest-frequency longitudinal optical mode which yields the largest electron-phonon coupling matrix elements has a frequency of around 110 meV, i.e., much larger than $k_B T$ at room temperature. This makes scattering of room-temperature thermalized electrons by phonon emission unlikely because most thermalized electrons do not have such high energies, while scattering by phonon absorption is also unlikely at room temperature because the Bose-Einstein occupation of these phonon modes is low (1%). As a result, piezoelectric electron-phonon interaction arising from the acoustic phonons plays a dominant role to electron scattering at room temperature.[19] This makes the inclusion of quadrupole corrections crucial for an accurate interpolation of the long-range component of the acoustic electron-phonon matrix elements (Section S3). We find that the inclusion of quadrupole corrections increases the room-temperature electron mobilities in AlN by 20-30% and is consequently important for accurate mobility calculations.

Next, we calculate the total mobility including the combined effects of both phonon and ionized-impurity scattering[46] to examine the dependence on electron density and ionized-impurity concentration. The total electron drift mobility is shown in Figure 2a as a function of ionized impurity concentration, under the two conditions of total ionization and partial ionization. For the partial ionization condition, we set the electron concentration to $10^{15}$ cm$^{-3}$, which represents the upper order of magnitude of typical electron concentrations reported in experimental AlN samples with high electron mobility values. The partial ionization condition represents the typical case of donor dopants in AlN, e.g., Si$_{Al}$, Ge$_{Al}$, or O$_N$, which form DX$^-$ centers[7] that not only reduce the free electron concentration, but also act as charge centers that contribute to scattering. For this reason, we do not consider the effect of scattering by neutral donors in this work and consider all donor ions to be charged either positively or negatively. For low ionized-impurity concentrations, phonon scattering dominates and we calculate high drift mobilities above 800 cm²/V·s and 600 cm²/V·s along the in-plane and out-of-plane directions, respectively, i.e., similar to the phonon-limited mobility values in the previous section. For ionized impurity concentration higher than $10^{16}$ cm$^{-3}$, ionized-impurity scattering suppresses the electron mobility, while at concentrations higher than $5\times10^{17}$ cm$^{-3}$ ionized impurities become the dominant source of electron scattering. At high doping concentrations, the electron mobility decreases to about 75 cm²/V·s in the case of total ionization and 5 cm²/V·s in the case of partial ionization. The partial-ionization condition yields lower mobility because free-carrier screening of the ionized-impurity scattering is reduced. We fit the ionized-impurity-concentration dependence of the total mobilities according to the following empirical model developed by Caughey and Thomas:[47]

$$\mu_{\text{total}}(n) = \mu_{\min} + \frac{\mu_{\max} - \mu_{\min}}{1 + \left(n/n_{\text{ref}}\right)^{\beta}}. \quad (2)$$

Here, $\mu_{\max}$ represents the phonon-limited mobility, $\mu_{\min}$ represents the lower limit of mobility when ionized impurity scattering is dominant, $\beta$ characterizes the density dependence of the mobility between the two limits, and $n_{\text{ref}}$ is the characteristic doping concentration for which the mobility value is halfway between the two limits. The first-principles data for the drift mobility are well described by the Caughey-Thomas model, and the fitted parameters are listed in Table SIII.

In addition to the drift mobility, we calculate the Hall mobilities to compare to values from experimental Hall measurements. We find that the two ionization conditions we examined describe well the range of available experimental data. The Hall mobilities, calculated with an additional magnetic field term in the Boltzmann transport equation, are slightly higher than the corresponding drift mobilities, with Hall factors ranging from 1.03 to 1.24. Figure 2b shows our calculated Hall mobilities plotted alongside experimentally measured mobilities. The experimental data points, which correspond to varying ionization ratios, fall either between the two ionization conditions that we consider or on the partial ionization condition. We note that in our calculations, we assume that the dopant concentration is equal to the ionized-impurity concentration (consisting of either positively charged donors or negatively charged DX⁻ centers), while the two concentrations are not necessarily equal in the experimental samples.

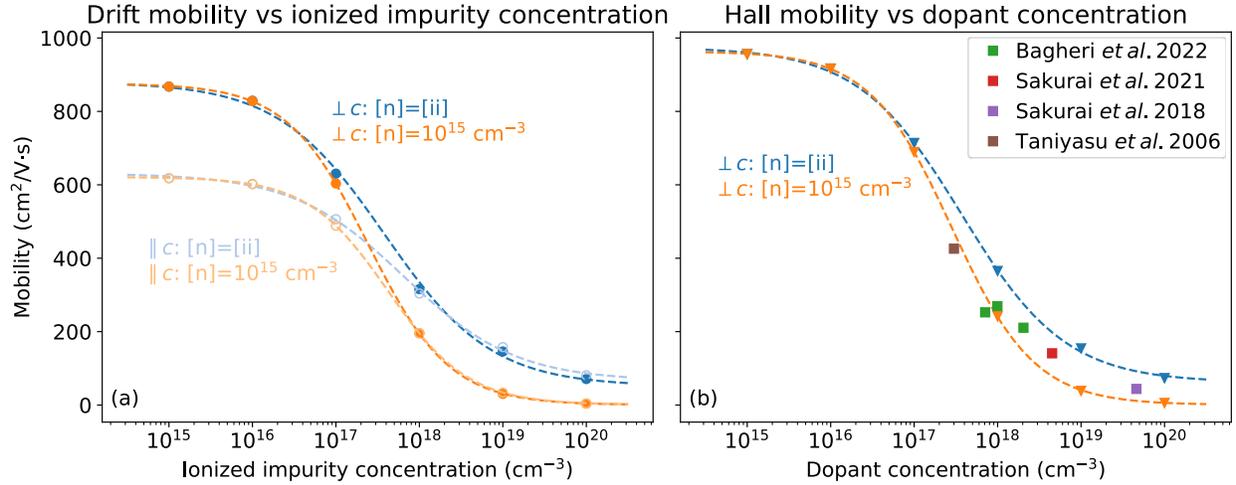

Figure 2 Total electron mobilities (including the effects of both phonon and ionized impurity scattering) of AlN as a function of concentration. Panel (a) shows the drift mobility along the in-plane and out-of-plane directions as a function of ionized impurity concentration. Two ionization conditions are shown: full ionization, under which the electron concentration and ionized impurity concentration are equal ($[n]=[ii]$), and partial ionization where the electron concentration is fixed to $10^{15}$ cm$^{-3}$ ($[n]=10^{15}$ cm$^{-3}$). Panel (b) shows the in-plane Hall mobilities under the same two ionization conditions, alongside experimentally measured Hall mobilities in Si-doped AlN from the works of Bagheri *et al.* (2022),[11] Sakurai *et al.* (2021),[13] Sakurai *et al.* (2018),[48] and Taniyasu *et al.* (2006).[12]

Our calculations show that drift mobilities as high as 867 cm$^2$/V·s and Hall mobilities as high as 956 cm$^2$/V·s are possible if the ionized impurity concentration is on the order of $10^{15}$ cm$^{-3}$ or less. This means the route to even higher mobilities than have been experimentally observed to date is the development of high-quality material samples with low dopant concentrations. Defects that commonly arise in n-type AlN are dislocations (especially if grown on lattice-mismatched substrates such as Al$_2$O$_3$), substitutional C$_N$, and Al vacancies, as well as the DX$^-$ centers formed by the dopants themselves.[6,11,12] These defects are detrimental to mobility in two ways: they both contribute additional scattering centers that impede electrons, and they also act as compensators that reduce the density of free electrons. The experimental works with high mobilities that we compare to all focused on reducing the defect density. Taniyasu *et al.*[12] found that at low doping concentrations, a high dislocation density severely impacts mobility because dislocations are not screened out by ionized impurities or free electrons. Their homoepitaxial AlN sample with a high mobility of 426 cm$^2$/V·s had a threading dislocation density of <$10^6$ cm$^{-2}$, i.e., three orders of magnitude lower than their heteroepitaxial sample with a mobility of 242 cm$^2$/V·s. The studies by Sakurai *et al.*[13,48] use high-temperature annealing to reduce the dislocation density and improve crystal quality. The work by Bagheri *et al.*[11] optimized growth conditions (temperature, precursor flow, and quasi-Fermi level) and used AlN substrates to minimize the formation of C$_N$ and dislocations, and achieve high mobilities at low doping concentrations. While these works represent significant progress in reaching higher mobilities through improved crystal quality, our theoretical results for the upper bounds of electron mobility in AlN demonstrate that there is still room for further reduction of the compensating defects and higher dopant ionization ratios to achieve even higher mobilities.

In conclusion, we calculate the electron mobility of AlN as a function of temperature, doping, and crystallographic orientation from first principles to identify the dominant scattering mechanisms and to determine the theoretical upper limit of transport in this ultra-wide band gap material. We find that long-range piezoelectric electron-phonon coupling due to the acoustic phonon modes dominates scattering in the phonon-limited mobility. We also calculate the total mobility accounting for both phonon and ionized-impurity scattering, assuming both full and partial donor ionization conditions. Our calculated values agree with the range of experimental mobility data from samples with varying ionization ratios. Ionized-impurity scattering substantially decreases the electron mobility at dopant concentrations above $10^{16}$ cm$^{-3}$. Our results demonstrate that even higher mobility values can be achieved in high-quality material samples, which suppress compensation and scattering by detrimental defects, and which allow doping with lower dopant concentrations.

**Supplementary Material**

See the supplementary material for more details on the calculation parameters and convergence, a comparison of the full band structure and effective masses calculated with the LDA functional and with $G_0W_0$, the quadrupole tensor, and the fitted mobility parameters.

**Acknowledgements**

The work is supported as part of the Computational Materials Sciences Program funded by the U.S. Department of Energy, Office of Science, Basic Energy Sciences under Award No. DESC0020129. Computational resources were provided by the National Energy Research Scientific Computing (NERSC) Center, a DOE Office of Science User Facility, under Contract No. DEAC02–05CH11231. A.W. is supported by the Department of Defense (DoD) through the National Defense Science & Engineering Graduate (NDSEG) Fellowship Program.

**Author Declarations**

**Conflict of Interest**

The authors have no conflicts to disclose.

**Data Availability**

The data that support the findings of this study are available from the corresponding author upon reasonable request.

# Electron mobility in AlN from first principles: supplementary material

Amanda Wang[1], Nick Pant[1,2], Woncheol Lee[3], Jie-Cheng Chen[4,5], Feliciano Giustino[4,5], Emmanouil Kioupakis[1]

[1] Department of Materials Science and Engineering, University of Michigan, Ann Arbor, Michigan 48109, USA

[2] Applied Physics Program, University of Michigan, Ann Arbor, Michigan 48109, USA

[3] Department of Electrical and Computer Engineering, University of Michigan, Ann Arbor, Michigan 48109, USA

[4] Oden Institute for Computational Engineering and Sciences, The University of Texas at Austin, Austin, Texas 78712, USA

[5] Department of Physics, The University of Texas at Austin, Austin, Texas 78712, USA

## S1 Calculation details

The DFT, DFPT, BerkeleyGW, and EPW calculations are run with norm-conserving pseudopotentials generated the Fritz-Haber code using the Troullier-Martins pseudization.[1,2] The ABINIT calculation to find the quadrupole tensor is run with norm-conserving pseudopotentials without non-linear core corrections from PseudoDojo.[3] The DFT and DFPT calculations are run with a plane-wave cutoff of 70 Ry and the SCF charge density is converged on a 10×10×10 Brillouin zone (BZ) sampling grid. The BerkeleyGW calculations are run with a dielectric cutoff of 20 Ry and 200 bands in the summation, including bands up to 11 Ry above the valence band maximum, to calculate the quasiparticle energies on an 8×8×8 BZ grid.

The convergence of the mobility values is tested with respect to both the fine BZ sampling and the energy window which determines which states near the edge of the conduction band minimum are considered. The mobility values are converged within 5% at a fine BZ sampling grid of 144×144×96 when compared to a 128×128×80 grid. Mobility calculations at lower carrier concentrations and lower temperatures need narrower energy windows compared to higher carrier concentrations or higher temperatures due to lower Fermi occupations. The mobility values are converged within 5% with respect to the energy window for the various carrier concentration and temperature conditions. Mobility at the lowest carrier concentration, $10^{15}$ cm$^{-3}$, is calculated with an energy window of 250 meV and mobility at the highest carrier concentration, $10^{20}$ cm$^{-3}$, is calculated with an energy of window of 325 meV. An energy window of 350 meV is used for the temperature-dependent mobilities to capture all the relevant states for the highest temperature in our calculations (600 K).

We calculate carrier mobility $\mu$ using the linear response of the carrier occupations $f_{nk}$ with respect to an applied electric field:

$$\mu_{\alpha\beta} = \frac{-1}{V_{uc}n_c} \sum_n \int \frac{d\mathbf{k}}{\Omega_{BZ}} v_{nk\alpha} \partial_{E_\beta} f_{nk}, \qquad (S1)$$

where $\alpha$ and $\beta$ are Cartesian directions, $\mathbf{v}_{nk}$ is the velocity of the state at band index $n$ and crystal momentum $\mathbf{k}$, and the linear response coefficient is defined as $\partial_{E_\beta} f_{nk} \equiv (\partial f_{nk}/\partial E_\beta)|_{E=0}$. The linear response coefficients are calculated by iteratively solving the linearized Boltzmann transport equation:

$$\partial_{E_\beta} f_{nk} = ev_{nk} \frac{\partial f_{nk}^0}{\partial \varepsilon_{nk}} \tau_{nk} + \tau_{nk} \sum_m \int \frac{d\mathbf{q}}{\Omega_{BZ}} \Gamma_{mk+q \to nk} \partial_{E_\beta} f_{mk+q}. \qquad (S2)$$

$f_{nk}^0$ is the equilibrium Fermi-Dirac occupation and $\varepsilon_{nk}$ is the energy of the $n\mathbf{k}$-th state. The integrals over $\mathbf{k}$ and $\mathbf{q}$ are computed as sums over the fine BZ sampling grid and only states within the energy window are considered in the sum. $\tau_{nk}$ is the state-dependent total scattering lifetime and its inverse is calculated as the sum over the partial transition rates $\Gamma_{nk \to mk+q}$:

$$\tau_{nk}^{-1} = \sum_m \int \frac{d\mathbf{q}}{\Omega_{BZ}} \Gamma_{nk \to mk+q}. \qquad (S3)$$

The phonon-limited mobilities are calculated using partial transition rates from electron-phonon scattering, while the total mobilities are calculated using the sum of the partial transition rates from both electron-phonon scattering and electron-ionized impurity scattering ($\Gamma_{nk \to mk+q} = \Gamma_{nk \to mk+q}^{ph} + \Gamma_{nk \to mk+q}^{ii}$).

The partial transition rate due to electron-phonon scattering is calculated using the interpolated electron-phonon matrix elements $g_{mn\nu}(\mathbf{k}, \mathbf{q})$, energy-conserving delta functions approximated as Gaussians, and the Fermi-Dirac and Bose-Einstein occupation factors $f_{nk}^0$ and $n_{q\nu}$:[4]

$$\Gamma_{nk \to mk+q}^{ph} = \sum_\nu |g_{mn\nu}(\mathbf{k}, \mathbf{q})|^2 \times [\delta(\varepsilon_{nk} - \varepsilon_{mk+q} + \hbar\omega_{q\nu})(1 + n_{q\nu} - f_{nk}^0)$$
$$+ (\varepsilon_{nk} - \varepsilon_{mk+q} - \hbar\omega_{q\nu})(n_{q\nu} + f_{nk}^0)]. \qquad (S4)$$

Here one can see the dependence of the scattering rate on both the electron-phonon matrix element and on the Bose-Einstein occupation of the phonons. The partial transition rate due to electron-ionized impurity scattering is calculated using an ensemble average of randomly distributed point charges.[5]

To calculate Hall mobility, or the response of the carrier drift velocity in the presence of both an electric and a magnetic field, the linearized BTE is augmented with a magnetic field term:

$$\left[1 - \frac{e}{\hbar}\tau_{nk}(v_{nk} \times B) \cdot \nabla_{\mathbf{k}}\right] \partial_{E_\beta} f_{nk} = ev_{nk}\frac{\partial f_{nk}^0}{\partial \varepsilon_{nk}}\tau_{nk} + \tau_{nk}\sum_{mq}\Gamma_{mk+q \to nk}\, \partial_{E_\beta} f_{mk+q}. \qquad (S5)$$

A small, finite magnetic field of 10⁻¹⁰ T is applied in the calculation and the gradient calculated with finite differences. The linear response coefficients from the self-consistent solution of this equation are used to calculate the Hall mobilities which we compare to experimentally measured mobilities.

## S2 Band structure and effective masses

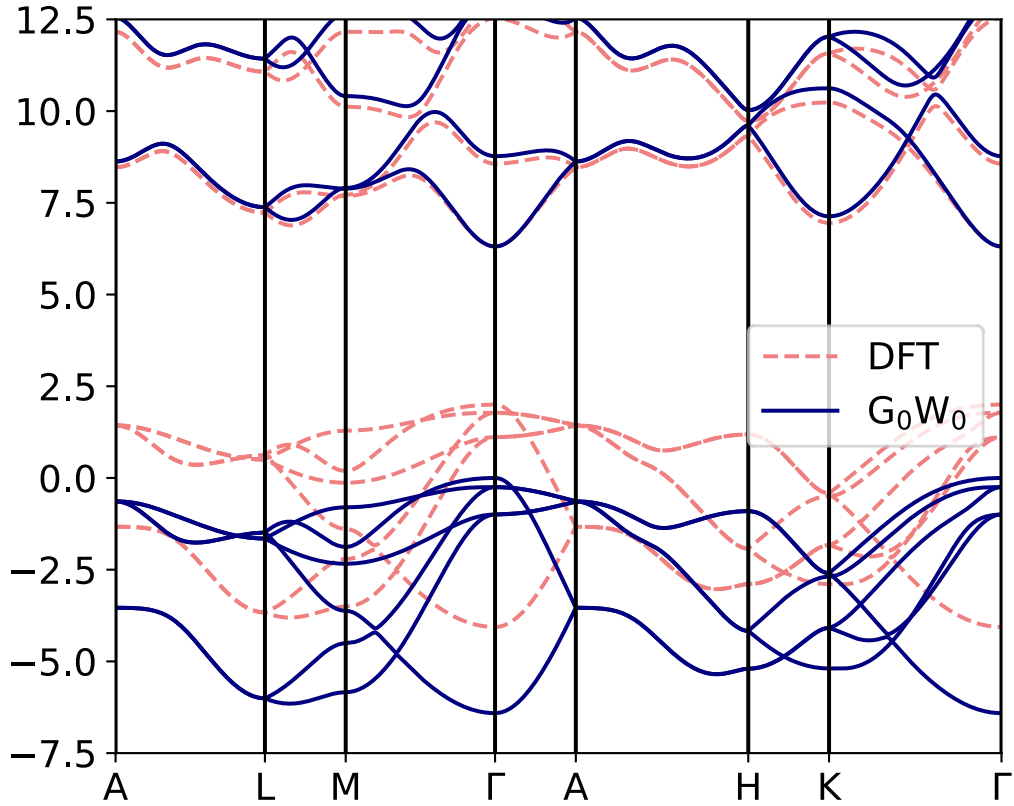

Figure S1 Band structure of AlN along high-symmetry paths of the Brillouin zone. The DFT eigenvalues are shown with the dashed pink line while the $G_0W_0$ quasiparticle energies are shown in solid blue.

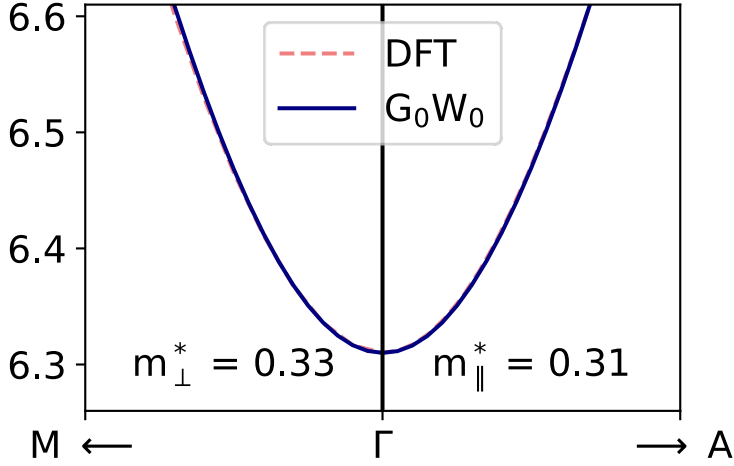

Figure S2 Bottom of the conduction band along the in-plane and out-of-plane directions with the corresponding effective masses fit to the hyperbolic model. The effective masses are fit to the DFT eigenvalues shown with the dashed pink line, which have good agreement at the bottom of the band with the $G_0W_0$ quasiparticle energies shown in solid blue.

## S3 Quadrupole tensor and piezoelectric coupling

Table SI The quadrupole tensor (in $e$ Bohr) of AlN calculated using ABINIT. The calculated values are used to improve the interpolation of the long-range component of the electron-phonon matrix elements by EPW. ($Q^{\beta\gamma}_{\kappa\alpha} = Q^{\gamma\beta}_{\kappa\alpha}$)

| κ | α | $Q^{11}$ | $Q^{22}$ | $Q^{33}$ | $Q^{23}$ | $Q^{13}$ | $Q^{12}$ |
|---|---|---|---|---|---|---|---|
| Al$_1$ | 1 | 0 | 0 | 0 | 0 | −1.80881 | −3.01714 |
|  | 2 | −3.01714 | 3.01714 | 0 | −1.80881 | 0 | 0 |
|  | 3 | −1.52705 | −1.52705 | 2.89497 | 0 | 0 | 0 |
| Al$_2$ | 1 | 0 | 0 | 0 | 0 | −1.80881 | 3.01714 |
|  | 2 | 3.01714 | −3.01714 | 0 | −1.80881 | 0 | 0 |
|  | 3 | −1.52705 | −1.52705 | 2.89497 | 0 | 0 | 0 |
| N$_1$ | 1 | 0 | 0 | 0 | 0 | 0.48227 | 0.46460 |
|  | 2 | 0.46460 | −0.46460 | 0 | 0.48227 | 0 | 0 |
|  | 3 | 0.12080 | 0.12080 | −1.19794 | 0 | 0 | 0 |
| N$_2$ | 1 | 0 | 0 | 0 | 0 | 0.48227 | −0.46460 |
|  | 2 | −0.46460 | 0.46460 | 0 | 0.48227 | 0 | 0 |
|  | 3 | 0.12080 | 0.12080 | −1.19794 | 0 | 0 | 0 |

To accurately interpolate the electron-phonon matrix elements when they contain long-range effects, they are separated into short-range and long-range parts where the short-range is interpolated using maximally localized Wannier functions and the long-range is calculated using the multipole expansion. We include the dipole and quadrupole terms in the calculation of the long-range component, using the Born effective charges calculated by DFPT to evaluate the dipole term and the quadrupole tensor shown above to evaluate the quadrupole term.[6] The relation between the quadrupole term of the electron-phonon matrix element and piezoelectric coupling can be seen in the phenomenological expression for piezoelectric coupling to acoustic phonons in ferroelectric materials:[7]

$$g_{mn}^{ac,piezo}(\boldsymbol{k}, \boldsymbol{q}) = -4\pi|e|\epsilon_L^{-1}(\boldsymbol{q} \to \boldsymbol{0})\langle u_{m\boldsymbol{k}+\boldsymbol{q}}|u_{n\boldsymbol{k}}\rangle \sum_{\alpha\beta\gamma} \frac{q_\beta}{q} \frac{q_\gamma}{q} e_{\beta\alpha\gamma} x_{\Gamma\alpha}^{ac} \quad (S6)$$

where $\epsilon_L^{-1}$ is the long-range dielectric function, $e_{\beta\alpha\gamma}$ is the piezoelectric tensor, and $x_{\Gamma\alpha}^{ac}$ is the contribution to atomic displacements from unit cell deformations by acoustic modes in the limit of $\boldsymbol{q} \to \boldsymbol{0}$. The piezoelectric tensor can be written as the sum of the quadrupole tensor over the atoms in the unit cell.[8] From this we can see that the piezoelectric coupling of electrons to acoustic phonon modes is a long-range effect that we use the quadrupole term to more accurately capture in our calculations.

## S4 Mobility parameters

We fit the temperature and carrier concentration dependence of the electron mobility to the analytical models described in the main text. Listed here are the fitted parameters of the models.

Table SII Fitted parameters for the temperature dependence of the phonon-limited drift mobilities.

|  | $\mu_{\text{low}}$ (cm²/V·s) | $\mu_{\text{high}}$ (cm²/V·s) | $T_{\text{low}}$ (K) | $T_{\text{high}}$ (K) |
|---|---|---|---|---|
| $\mu_{\text{ph}}(T) \perp c$ | 2415.09 | 18.65 | 96.39 | 1233.32 |
| $\mu_{\text{ph}}(T) \parallel c$ | 1191.77 | 17.45 | 83.35 | 1209.53 |

Table SIII Fitted parameters for the ionized impurity concentration dependence of the total drift and Hall mobility, based on the Caughey-Thomas model.

|  | $\mu_{\text{min}}$ (cm²/V·s) | $\mu_{\text{max}}$ (cm²/V·s) | $\beta$ | $n_{\text{ref}}$ (cm⁻³) |
|---|---|---|---|---|
| $\mu_{\text{total}}(ii) \parallel c : [n = ii]$ | 51.20 | 879.76 | 0.68 | 3.82×10¹⁷ |
| $\mu_{\text{total}}(ii) \perp c : [n = ii]$ | 65.85 | 631.00 | 0.65 | 6.66×10¹⁷ |
| $\mu_{\text{total}}(ii) \parallel c : [n = 10^{15}]$ | 0.38 | 875.50 | 0.89 | 2.44×10¹⁷ |
| $\mu_{\text{total}}(ii) \perp c : [n = 10^{15}]$ | 0.79 | 621.29 | 0.91 | 4.21×10¹⁷ |
| $\mu_{\text{total}}^{\text{Hall}}(ii) \parallel c : [n = ii]$ | 61.65 | 974.29 | 0.70 | 3.75×10¹⁷ |
| $\mu_{\text{total}}^{\text{Hall}}(ii) \perp c : [n = ii]$ | 61.95 | 663.55 | 0.71 | 9.33×10$^{17}$ |
| $\mu_{\text{total}}^{\text{Hall}}(ii) \parallel c : [n = 10^{15}]$ | 0.58 | 963.57 | 0.88 | 2.85×10$^{17}$ |
| $\mu_{\text{total}}^{\text{Hall}}(ii) \perp c : [n = 10^{15}]$ | 1.39 | 672.90 | 0.91 | 5.04×10$^{17}$ |